# CF$^2$-Net: Coarse-to-Fine Fusion Convolutional Network for Breast Ultrasound Image Segmentation

Zhenyuan Ning+, Ke Wang+, Shengzhou Zhong, Qianjin Feng, Yu Zhang*

*Abstract*—Breast ultrasound (BUS) image segmentation plays a crucial role in a computer-aided diagnosis system, which is regarded as a useful tool to help increase the accuracy of breast cancer diagnosis. Recently, many deep learning methods have been developed for segmentation of BUS image and show some advantages compared with conventional region-, model-, and traditional learning-based methods. However, previous deep learning methods typically use skip-connection to concatenate the encoder and decoder, which might not make full fusion of coarse-to-fine features from encoder and decoder. Since the structure and edge of lesion in BUS image are common blurred, these would make it difficult to learn the discriminant information of structure and edge, and reduce the performance. To this end, we propose and evaluate a coarse-to-fine fusion convolutional network (CF$^2$-Net) based on a novel feature integration strategy (forming an "E"-like type) for BUS image segmentation. To enhance contour and provide structural information, we concatenate a super-pixel image and the original image as the input of CF$^2$-Net. Meanwhile, to highlight the differences in the lesion regions with variable sizes and relieve the imbalance issue, we further design a weighted-balanced loss function to train the CF$^2$-Net effectively. The proposed CF$^2$-Net was evaluated on an open dataset by using four-fold cross validation. The results of the experiment demonstrate that the CF$^2$-Net obtains state-of-the-art performance when compared with other deep learning-based methods.

*Index Terms*—Coarse-to-fine fusion, convolutional neural network, deep learning, breast ultrasound image segmentation

## I. INTRODUCTION

BREAST cancer, as the most common type of cancer among females, is one of the leading causes of death for females worldwide, and its early diagnosis is crucial to reduce the mortality rate [1], [2]. Two screening modalities, namely, mammography and breast ultrasound (BUS), are popular imaging tools for the detection and diagnosis of breast cancer [1]. Although mammography is more commonly used than BUS in clinical practice, it suffers from some limitations: i) high false positive rate and ii) insensitivity to dense breast tissues [3], [4]. Compared with mammography, BUS has the advantages of being radiation-free, having high sensitivity, easy accessibility, and low cost [5]. Therefore, BUS has become an alternative imaging tool for breast cancer. Recently, many computer-aided diagnosis (CAD) systems using BUS images have been proposed for the early detection and diagnosis of breast cancer [6], [7], [8]. Generally, a CAD system based on BUS mainly refers to image segmentation, feature extraction, and model construction [9]. Image segmentation is a key step in the CAD system, as segmentation can directly affect the accuracy and robustness of diagnosis [10]. However, manual image segmentation requires the skill and experience of a radiologist, and even well-trained radiologist may have a high inter-observer variation rate [1], [3]. Although many segmentation approaches have been proposed to replace manual segmentation, it still remains challenging due to the low image quality caused by speckle noise, low contrast, and artifacts [11], [12]. In addition, since the lesions in BUS image have the characteristics of various size, shape, and blurred edge (as shown in Fig.1), which would increase the difficulty of segmentation [5], [13].

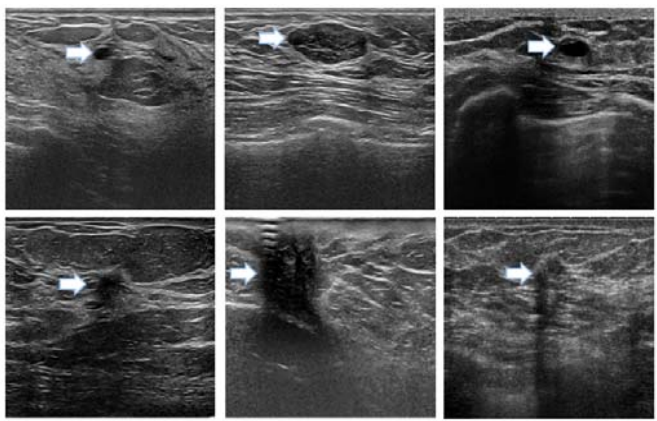

Fig.1. Some examples of BUS images used in our work. The top row shows the images of begin lesions, and bottom row shows the images of malignant lesions. As the figure shown, the regions have various size and shape, and blurred edge regardless of begin or malignant lesions.

Numerous approaches have been developed for BUS image segmentation, including region-, model-, and learning-based methods [3]. Region-based methods [14], [15], [16], [17] perform image segmentation based on the similarity of neighborhood pixels or regions, and prior constraints, such as shape, appearance, and spatial location of the lesion regions. Model-based methods [18], [19], [20], [21], [22] achieve broad application in BUS image segmentation by constructing models

Z. Ning, K. Wang, S. Zhong, Q. Feng, and *Y. Zhang is with the School of Biomedical Engineering, Southern Medical University, Guangzhou, Guangdong, 510515, China (*E-mail: yuzhang@smu.edu.cn). Z. Ning and K. Wang contributed equally to this work.

using unique prior/posterior-based energy functions and optimization algorithms, including the graph-based model [19], the active contour model [20], and the Markov random field model [21]. However, region- and model-based methods usually require the similarity rules, prior constraints, a series of complicated preprocessing operators, and accurate initialization points, which depend on the knowledge and experience of researchers. In addition, these methods are sensitive to noise and inclined to over-segmentation. Learning-based methods aim to learn certain potential functions that can map the original image space to the target space. Conventionally, learning-based methods [23], [24] consider a segmentation task as a pixel-based classification problem that can be solved by various machine learning approaches. The segmentation performance of traditional learning-based common depends on discriminative feature extraction [25]. Moreover, feature extraction and classifier construction are standalone, thereby the segmentation performance is suboptimal [26].

Recently, deep learning methods, especially convolutional neural networks (CNNs), have been successfully applied to BUS image segmentation [25] and show some advantages by integrating high-level nonlinear feature extraction with model construction in an end-to-end manner [27], [28], [29]. The existing CNN-based methods in BUS image segmentation can be divided into two subnets, namely, encoder and decoder [30], [31]. For example, Bian et al. [32] proposed an encoding-decoding network to address the different anatomical layer paring problem in BUS image segmentation. Yuan et al. [27] used CNN to segment BUS images into four major tissues, i.e., skin, fibro glandular, mass, and fatty tissue. Hu et al. [28] developed an BUS image segmentation model by combining a dilatational fully convolutional network with a phase-based activated contour model. Xing et al. [29] proposed a new semi-pixel-wise cycle generative adversarial network for segmenting lesion in the ultrasound image. However, these CNN-based methods typically use skip-connection to concatenate the encoder and decoder, which might not make full use fusion of coarse-to-fine features from encoder and decoder. The structure and edge of lesion in BUS image is common blurred, which would make it difficult to learn the discriminant information of structure and edge, and reduce the performance.

To address the aforementioned issues, we propose a novel coarse-to-fine fusion convolutional network (CF$^2$-Net) for BUS image segmentation. Specifically, U-Net [30] is used as the backbone network of the CF$^2$-Net to generate coarse-to-fine feature maps. Instead of using skip-connection, we propose a novel fusion stream path (FSP) to hierarchically integrate coarse-to-fine information from the encoding and decoding path of the backbone network. The FSP consists of four modules, and each module contains four units, including atrous spatial pyramid pooling (ASPP) unit, cascade feature fusion (CFF) unit, edge constraint (EC) unit and a tiny U-Net unit. Firstly, the ASPP unit is used to obtains abundant receptive fields for simultaneously capturing characteristics of target region with different sizes. After that, a cascade feature fusion (CFF) unit fuses coarse-to-fine and low-to-high level information by contextual transmission strategy. To remit the blurred edge issue, an edge constraint (EC) unit is embedded to make the network with the ability of capturing edge information. In addition, we further concatenate a super-pixel image and the original image as inputs to enhance contour and provide additional structure information. In the end, the CF$^2$-Net is trained using the weighted-balanced loss function with self-weight for each image, which can relieve the imbalance issue between the target region and the background, and adapt to the challenge of the various sizes of the lesion region. The main contributions of this work are summarized as follows:

- To the best of our knowledge, we first propose a novel additional stream for the segmentation network (forming a n "E"-like type), which hierarchically integrates coarse-to-fine features from encoder and decoder. Moreover, it can capture additional edge information constrained by an EC unit. The "E"-like type of network can also be used in other segmentation tasks.
- Considering the characteristics of BUS images, namely, blurred edge and structure, the super-pixel image, as prior knowledge, is concatenated with the original image for the inputs of network. It would encourage the network to learn additional structure and edge information. The experiment shows that the super-pixel strategy used as the input of the network is superior to the network that used it for post-processing.
- To remit the imbalance between the target region and the background and to adapt the characteristics of various tumor sizes, we employ a weighted-balanced loss function with self-weight for each image to train the CF$^2$-Net effectively.

The rest of this paper is organized as follows. In Section 2, the dataset used in this study is introduced, and the proposed CF$^2$-Net is described in detail. In Section 3, the experimental settings and results are exhibited and analyzed. The discussion and conclusion are presented in Section 4.

## II.  METHOD

In this section, we first introduce the dataset used in our study. Subsequently, we provide a detailed description of the proposed method for BUS image segmentation.

### *A. Datasets*

The dataset used to evaluate the proposed method is a public BUS image database available at goo.gl/SJmoti [1], which was collected from the UDIAT Diagnostic Center of the Parc Taulí Corporation, Sabadell (Spain) with a Siemens ACUSON Sequoia C512 SYSTEM 17L5 HD array transducer (8.5 MHZ). The dataset consisted of 163 images and each image contained one or more lesions with various sizes and shapes. The dataset contained 110 images with benign lesions and 53 images with cancerous masses. Among the benign images, 65 were unspecified cysts, 39 were fibro adenomas, and 6 were other types of benign lesions. Out of the malignant images, 40 were ductal carcinomas, 4 were invasive ductal carcinomas in situ, 2 were invasive lobular carcinomas, and 7 were other unspecified malignant lesions. The lesions in the dataset in this dataset were delineated by experienced radiologists. We first resized all the images into a fixed size of 255×255. Considering that the EC unit is an important component in our CF$^2$-Net, we then used a dilation algorithm to expand five pixels around the ground truth to extract the edge of lesion to train the network.

*B. Network Architecture*

As shown in Fig.2(a), the proposed CF$^2$-Net mainly contains two subnets, namely, the backbone U-Net (gray path) and the FSP (green path). Specifically, the U-Net is used as the backbone network, which contains an encoding and a decoding stream to generate coarse-to-fine features for the FSP. The feature maps from the encoding stream are shallow. In contrast, the feature maps from the decoding stream are deep. The most of current methods based on the classical networks, such as the U-Net and the fully convolutional network, usually utilize skip connection strategy to transmit shallow feature maps to the decoding stream and combine them for compensating the information loss during pooling and upsampling. However, the skip connection strategy is forward and unidirectional, which can neither adequately integrate the shallow-to-deep, coarse-to-fine information nor selectively eliminate redundancy information from the encode stream and help the decoding stream construct the segmented map for BUS images that contain high noise and low contraction. In our work, the FSP of CF$^2$-Net, as shown in the middle of Fig.2(a), consists of four modules that can integrate the feature maps generated by the corresponding symmetrical encoding and decoding streams in a hierarchical fusion manner. Specially, each FSP module contains an ASPP unit, a CFF unit, an EC unit, and a tiny U-Net unit. Fig.2(b) shows the detailed structure of each module. Finally, the ultimate outputs of the CF$^2$-Net are from the FSP, the backbone U-Net and the edge constraint.

*1) Input of the CF$^2$-Net*. For BUS image segmentation, capturing precise edge information is tough owing to blurring boundary, low contract and seriously noise and shadow. However, the convolutional neural network aims to extract similarity of texture and region boundary in early layers and the similarity in the network is difficult to quantify directly. Super-pixel algorithm [33] can eliminate the noise and enhance the boundary by generating irregular patches according to the similarity of neighborhood pixels. Instead of utilizing super-pixel algorithm to optimize the output of network [34], we concatenate a super-pixel image and the original image as multi-channel inputs for our proposed CF2-Net. Considering the simplicity and efficiency of the simple linear iterative clustering (SLIC) algorithm, we use this algorithm to obtain the super-pixel image as a complementary input channel to aid in capturing the neighborhood pixel similarity during the training process. On the one hand, SLIC can generate compact patches, maintain contour integrity, provide additional details for the network. On the other hand, information from the original image channel can guide the network in fine-tuning the segmentation result. For each BUS image, we obtain 2,000 super-pixels by using the entropy rate-based algorithm [33].

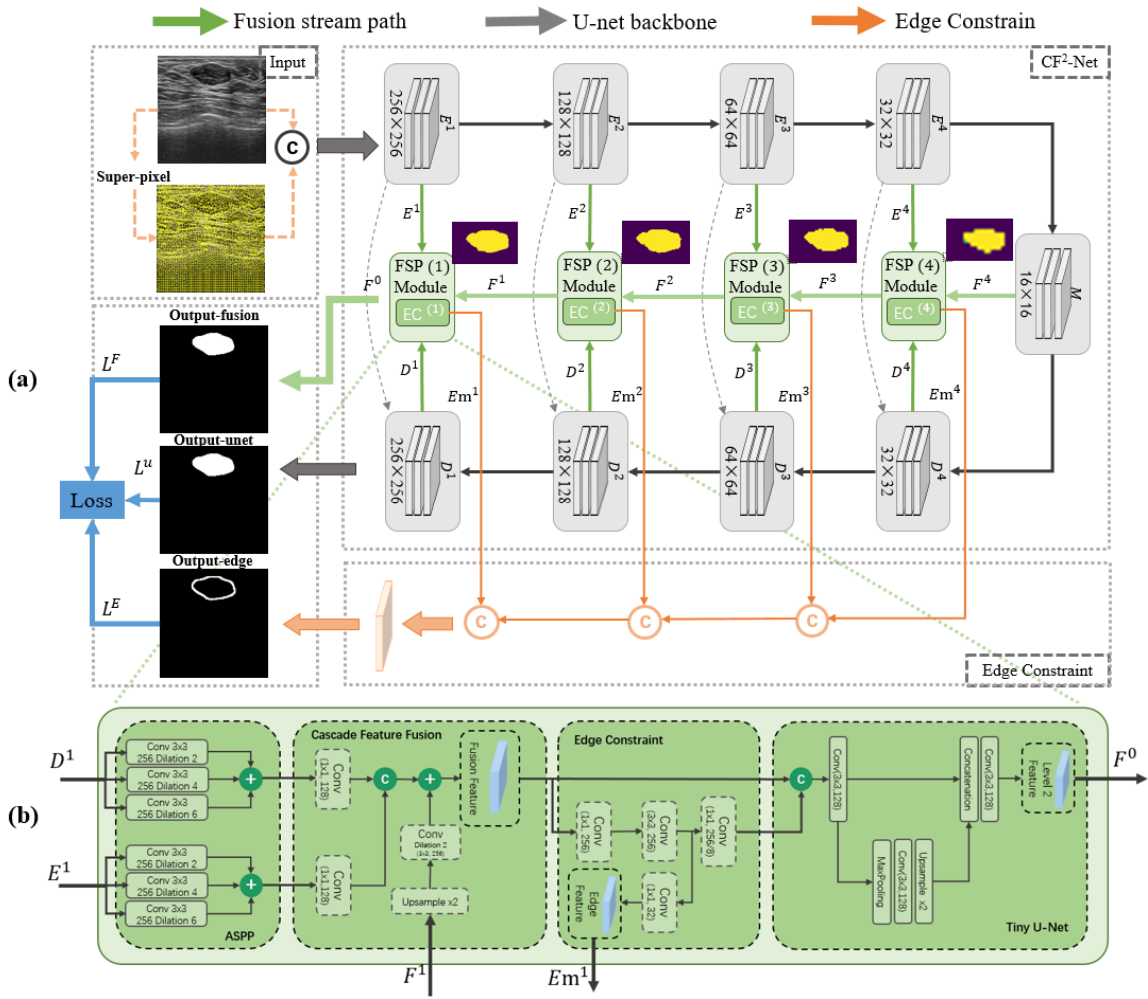

Fig.2. Overview of the proposed coarse-to-fine fusion convolutional network. (a): Our CF2-Net has two subnets, backbone U-Net and Fusion Stream Path (FSP) (green path). The backbone U-Net (gray path) extracts the information of low-high level from encoder and decoder. The FSP fuses the features using coarse-to-fine way. Adapting super-pixel algorithm and edge constraint to further improve the segmentation prediction. (b). The structure of FSP module. The module consists of ASPP, CF, EC units and a tiny U-Net.

*2) The backbone of the CF2-Net.* As illustrated in Fig.2(a), the U-Net is used as the backbone of CF2-Net to generate symmetric-scale feature maps, which consists of an encoding stream, a decoding stream, and a middle block (denoted as M). Specifically, the encoding steam is made up of four blocks (denoted as $E^i$, where $i \in \{1,2,3,4\}$ represents the i-th block). Moreover, each block $E^i$ involves three consecutive convolutional layers with the same kernel size of 3×3 and a stride of 1 to extract features under the i-th scale. Subsequently, a batch normalization (BN) layer is placed to relieve gradient vanishing and explosion problems, followed by a rectified linear unit (ReLU) activation layer to enhance nonlinearity. At the end of each block $E^i$, a "max" pooling layer with the size of 2×2 is used to reduce the feature dimensionality and retain spatial transformation invariance. Accordingly, the number of feature maps generated by each block $E^i$ is 64, 128, 256, and 512. The decoding stream shares a similar structure and uses an upsampling layer to take the place of the pooling layer in each block $D^i$ to interpolate the missed pixel and recover the size of the feature map stepwise. To retain the symmetric characteristic, the number of feature maps extracted by each block $D^i$ is also set to 64, 128, 256, and 512. The middle block M only discards the pooling or upsampling layer and generates 1024 feature maps.

*3) FSP.* The FSP of CF2-Net is a novel aggregation network that consists of four FSP modules that are armed with ASPP, CFF, EC, and tiny U-Net units to integrate coarse-to-fine information for BUS image segmentation.

● *ASPP unit.* Considering that the shape and size of the lesion region in the BUS image are commonly irregular and various, we first construct an ASPP unit to obtain abundant receptive fields to capture the characteristics of the lesion region with various sizes simultaneously. As illustrated in Fig.2(b), leveraging the feature maps generated by each encoding block $E^i$ and $D^i$ under the i-th scale, two convolutional groups are independently placed to further extract the features for block $E^i$ and $D^i$, respectively. And each convolutional group includes three parallel dilatational convolutional layers with the same kernel size of 3×3 and dilation rate of $\{2,4,6\}$. Subsequently, a ReLU activation layer is applied to enhance the nonlinearity of the network, followed by an aggregative operation to integrate the feature maps extracted by three dilatational convolutional layers.

● *CFF unit.* Currently, most works typically use skip-connection to concatenate the encoder and decoder for compensating the information loss during the pooling operation. We utilize the diversity of the features from encoder and decoder and extract feature to eliminate the redundant information and complement lost information To this end, we develop a CFF unit to integrate the symmetric feature maps under the current scale ($D^i$ and $E^i$) with the feature maps generated by the last tiny U-Net ($L^{i-1}$) under the (i-1)-th scale, which can efficiently fuse shallow-to-deep and coarse-to-fine information in stepwise contextual manner. As shown in Fig.2(b), an upsampling layer is first applied to $L^{i-1}$ for obtaining feature maps with the size of the i-th scale by bilinear interpolation algorithm, a dilatational convolutional layer with kernel size of 3×3 and rate of 2, and a BN layer are used to further extract and refine the features, respectively. The upsampling layer followed by dilatational convolutional layer can harvest the same receptive field and cost less computation than the deconvolutional layer. For the feature maps under the current scale ($D^i$ and $E^i$), we first use two 1×1 convolutional layers to extract semantic features and reduce the dimension of the feature for $D^i$ and $E^i$, respectively. Then, a concatenated layer integrates the feature maps of encoding $E^i$ and decoding $D^i$. Subsequently, an aggregative layer with ReLU activation combines these features from the concatenated layer and the BN layer to generate the fused features as the input of the EC unit.

● *EC unit.* The edge context can help restrain the integrality and the uniqueness of the region, especially when the BUS image contains various irregular lesions with low contrast. However, most works on BUS image segmentation ignore this issue. Therefore, we embed an EC unit in each FSP module to address the problem. As illustrated in Fig.2(b), two consecutive convolutional layers with kernel sizes of 1×1 and 3×3, respectively, are used to extract the edge information from fused features generated by the CFF unit. Then, a 1×1 convolutional layer further extracts high-level edge semantic features, which concatenates with the fused feature (from the CFF unit) as the input of the tiny U-Net. Meanwhile, another 1×1 convolutional layer is placed to obtain 32 feature maps (denoted as $Em^i$) for edge prediction at the current scale. After getting under four different scales, an upsampling layer is placed to acquire a feature map with the same size of 256×256. The final prediction of edge is computed as

$$E^{prediction} = \sigma(f_k[C\{up(Em^1,1),up(Em^2,2),up(Em^3,4),up(Em^4,8)\}]) \quad (1)$$

where $Em^i (i \in \{1,2,3,4\})$ are the feature maps generated by four EC units. $up(\cdot)$ defines the upsampling layer, and its second parameter is the sample rate. $C(\cdot)$ represents the concatenated layer. $f_k(\cdot)$ is the convolutional layer, and $\sigma(\cdot)$ denotes the sigmoid activation function.

● *Tiny U-Net unit.* To capture global information, a tiny U-Net unit is applied to extract high-level features based on concatenated features from the CFF and EC units. Specifically, the tiny U-Net unit consists of three consecutive convolutional layers with the same kernel size of 3×3.

Moreover, the second convolutional layer is placed between the pooling layer and the upsampling layer to extract global features. Finally, a concatenated layer and a 3×3 convolutional layer combine and fuse the global features with the original features, respectively. Notably, a BN layer and a ReLU activation layer are employed behind each convolutional layer.

*C. Weighted-balanced loss function*

For BUS image segmentation, class imbalance, which is defined as the tremendous differences in the number of pixels between the lesion region and the background, still remains a challenge due to lesion regions with various sizes. To this end, we design a weighted-balanced loss function to train the proposed network efficiently, which is based on a commonly used hybrid loss function that contains dice and binary cross entropy loss [35], [36], [37]. As illustrated in Fig.2(a), the final loss ( $\mathcal{L}$ ) of the scheme consists of the fusion path loss ( $L^F$ ) of the FSP, the auxiliary loss ( $L^U$ ) of the backbone U-Net, and the edge loss ( $L^E$ ) of the EC unit, and is computed as follows:

$$\mathcal{L} = \lambda_1 L^F + \lambda_2 L^U + \lambda_3 L^E \quad (2)$$

where $\lambda_1$, $\lambda_2$, and $\lambda_3$ are the weight coefficients and are experimentally assigned as 1, 1, and 0.1, respectively. A weighted-balanced parameter is embedded in $L^F$, $L^U$, and $L^E$ to address the class-imbalance issue in the BUS image segmentation. Specifically, $Y^s = \{y_i^s\}_{i=1}^N$ and $Y^e = \{y_i^e\}_{i=1}^N$ denote the ground truth of the lesion region and the edge, respectively, where $y_i^s = 1$ ( $y_i^s = 0$ ) and $y_i^e = 1$ ( $y_i^e = 0$ ) represent the i-th pixel that belongs to the tumor region (background) and the edge region (non-edge region), respectively; N denotes the number of pixels. $P^s = \{p_i^s\}_{i=1}^N$ and $P^e = \{p_i^e\}_{i=1}^N$ represent the predicted tumor region probability map and the predicted edge region probability map, respectively, where $p_i^s, p_i^e \in [0,1]$. $L^F$ and $L^U$ are expressed as follows:

$$L^* = \mu_1 L_d^* + \mu_2 L_c^* \quad (3)$$

where * denotes F or U. $\mu_1$ and $\mu_2$ are the weight coefficients. $L_d^*$ and $L_c^*$ are the dice loss with weighted-balanced parameter and binary cross entropy loss, respectively, and are computed as follows:

$$L_d^* = 1 - w_s^* \frac{\sum_{n=1}^{N_1^s} p_n^s y_n^s}{\sum_{n=1}^{N_1^s} (p_n^s + y_n^s)} - (1 - w_s^*) \frac{\sum_{m=1}^{N_0^s} (1-p_m^s)(1-y_m^s)}{\sum_{m=1}^{N_0^s} (2 - p_m^s - y_m^s)} \quad (4)$$

$$L_c^* = \frac{1}{N} \sum_{n=1}^{N} (p_n^s \log(y_n^s) + (1-p_n^s) \log(1-y_n^s)) \quad (5)$$

where $w_s^* = N_1^s / (N_1^s + N_0^s)$ is the weighted-balanced parameter. $N_1^s$ and $N_0^s$ denote the pixel number of $y_i^s = 1$ and $y_i^s = 0$, respectively. Similarly, $L^E$ is computed as follows:

$$L^E = -w_e \sum_{n=1}^{N_1^e} p_n^e \log(y_n^e) - (1-w_e) \sum_{m=1}^{N_0^e} (1-p_m^e) \log(1-y_m^e) \quad (6)$$

where $w_e = N_1^e / (N_1^e + N_0^e)$ is the weighted-balanced parameter, and $N_1^e$ and $N_0^e$ denote the pixel numbers of $y_i^e = 1$ and $y_i^e = 0$, respectively.

## III. EXPERIMENTS AND ANALYSES

In this section, we first introduce the evaluation and implementation tool. Then the experimental results are shown, which contain the results of the ablation experiments to validate the effect of each unit in the CF2-Net and the results of the comparison experiments with several current state-of-the-art segmentation methods for BUS images.

*A. Evaluation*

In this work, the performance of the proposed CF2-Net is comprehensively evaluated by three metrics, i.e., dice similarity coefficient (DSC), sensitivity (SEN), and positive prediction value (PPV), which are defined as

$$DSC = \frac{2TP}{2TP + FP + FN} \quad (7)$$

$$SEN = \frac{TP}{FP + FN} \quad (8)$$

$$PPV = \frac{TP}{TP + FP} \quad (9)$$

where TP (true positive) and FP (false positive), respectively, denote the number of predicted tumor pixels inside and outside the ground truth of the tumor. FN (false negative) and TN (true negative) represent the number of predicted background pixels inside and outside the ground truth of the tumor, respectively. All experimental results are reported by averaging the performance of four-fold cross validation.

*B. Implementation tool*

The proposed method is implemented using Python 3.6 based on Keras package under the Ubuntu 14.04 system with the NVidia GeForce Titan graphics cards. We adopt an adaptive gradient algorithm (Adagrad) optimizer with momentum 0.9 to train the network. The mini-batch size in each epoch was set to 4, and the epoch is 500, and the learning rate is 0.0006. All the network parameters are initialized by "he_normal" in [38].

*C. Results*

We perform a series of experiments to evaluate each key module of the proposed CF2-Net and make extensive comparisons with other methods for BUS image segmentation. Notably, we use the U-Net as the baseline to validate our model by stepwise adding a new module.

*1) The influence of weighted-balanced loss function*

In this paper, a weighted-balanced loss function is designed to alleviate the class-imbalanced issue. To verify the effectiveness of the loss function, we compare it with commonly used loss function without weighted-balanced parameters by using the U-Net backbone. For the simplicity of description, we denote the U-Net trained by loss function with and without weighted-balanced parameter as U-NetW and U-Net, respectively. As presented in Table I, the U-NetW achieves better performance (average improvement approximately 1.971% for DSC, 2.847% for SEN, and 0.866% for PPV) than the U-Net. From Fig.3, we can observe that the weighted-balanced loss function can improve the segmentation results by filling the gaps and enhancing the edges. This improvement further proves that the weighted-balanced parameter has the ability to relieve the class-imbalance issue.

TABLE I
SEGMENTATION RESULTS FOR VALIDATING THE EFFECTIVENESS OF LOSS FUNCTION WITH WEIGHTED- BALANCED PARAMETER.

|  | DSC(%) | SEN(%) | PPV(%) |
|---|---|---|---|
| U-Net | 80.035±4.340 | 81.860±3.854 | 83.005±5.429 |
| U-NetW | **82.006±2.837** | **84.707±1.177** | **83.871±6.031** |

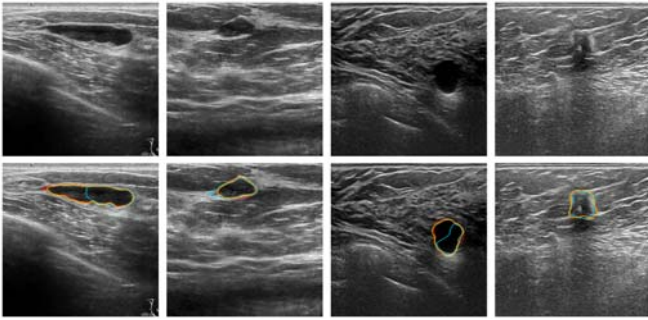

Fig.3. The visual comparison between the segmentation results of U-Net trained with commonly used loss function (blue) and trained weighted-balanced loss function with balanced parameter (yellow), and the ground truth is also plotted (red).

*2) The influence of incorporating the FSP module with U-Net*

● the CFF unit. As a main contribution of this work, the proposed CF$^2$-Net model extends the U-NetW with a combination of coarse-to-fine scale features from the decoding and encoding path via CFF units in FSP modules. To validate the effectiveness of the different units in the FSP module, we only use the CFF and the tiny U-Net units in the FSP module (denoted as CF$^2$-Net-C). The training procedure of the CF$^2$-Net-C and the U-NetW remains the same. The experimental results are listed in Table II. We can observe that the CFF unit can boost the performance of the U-NetW (improved 2.126%, 1.196%, and 2.573% for DSC, SEN, and PPV, respectively). The visual comparison between the segmentation results of the U-NetW and the CF$^2$-Net-C are shown in Fig.4. From Fig.4, we can observe that the CF$^2$-Net-C achieves better visual segmentation results than the U-Net by remitting under- and over-segmentation, which means that an efficient hierarchical fusion strategy for low-to-high level and coarse-to-fine scale features is beneficial to provide additional details to improve the performance of segmentation for BUS images.

TABLE II
SEGMENTATION RESULTS FOR PROVING THE FUSION STRATEGY FOR ENCODER AND DECODER IS USEFUL TO IMPROVE THE PERFORMANCE

|  | DSC(%) | SEN(%) | PPV(%) |
|---|---|---|---|
| U-NetW | 82.006±2.837 | 84.707±1.117 | 83.871±6.031 |
| CF2-Net-C | **84.132±2.841** | **85.903±2.253** | **86.544±4.162** |

● the ASPP and EC units. The characteristics of various shapes and sizes and the blurry edge of the region in a BUS image increases the complexity of segmentation, which is commonly ignored in most networks. To this end, we embed the ASPP and EC units in the FSP module. We perform a series of experiments to validate the effectiveness of the ASPP and EC units. Table III shows the experimental results, from which we can observe that both the ASPP and EC units can improve the performance of the segmentation (improved 0.789% and 1.813% for DSC and SEN, respectively). The PPV can be increased by only using the ASPP unit, as this unit may catch more details of regions with various sizes. However, jointly using ASPP and EC units results in the reduction of the PPV due to the lack of rich location information, which can be improved by multi-channel input. Fig.5 shows the visual comparison among the segmentation results of CF$^2$-Net-C, CF$^2$-Net-C with the ASPP unit and the joining of the ASPP and EC units. Evidently, the ASPP and EC units are beneficial to the segmentation of the region with various sizes and blurry edge.

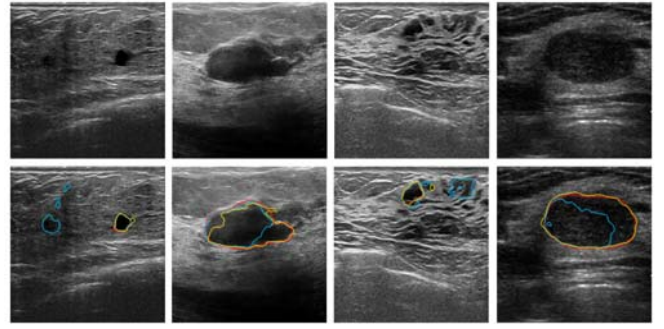

Fig.4. The visual comparison between the segmentation results of U-NetW (blue), CF2-Net-C (yellow), and the ground truth is also plotted (red).

TABLE III
SEGMENTATION RESULTS FOR TESTING THE EFFECTIVENESS OF ATROUS SPATIAL PYRAMID POOLING (ASPP) UNIT AND EDGE CONSTRAIN (EC) UNIT.

| ASPP | EC | DSC(%) | SEN(%) | PPV(%) |
|---|---|---|---|---|
| × | × | 84.132±2.841 | 85.903±2.253 | 86.544±4.162 |
| √ | × | 84.300±2.522 | 85.962±2.451 | **87.254±2.619** |
| √ | √ | **84.930±2.679** | **87.716±2.080** | 85.701±3.882 |

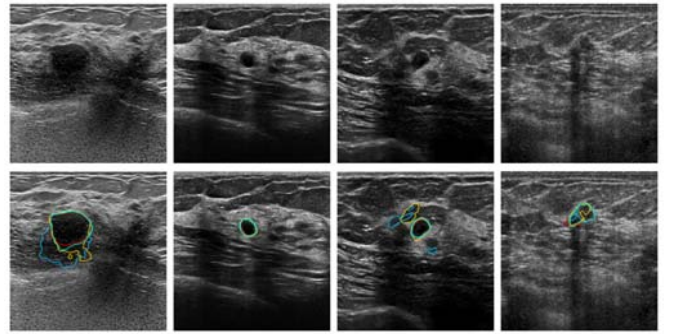

Fig.5. The visual comparison between the segmentation results of CF2-Net-C (blue), CF2-Net-C with ASPP unit (yellow), CF2-Net-C with ASPP and EC (green) units, and the ground truth is also plotted (red).

*3) The influence of super-pixel input*

To maintain contour integrity and provide information about textual similarity, we use super-pixel images as one of the inputs of CF$^2$-Net. After obtaining the image generated by the super-pixel algorithm, we concatenate it with the original image as multi-channel inputs. To assess the effectiveness of this strategy, we compare it with the model without super-pixel constraint. In Table IV, the super-pixel can improve the performance of the segmentation, especially in PPV (improved 0.623%, 0.495%, and 2.497% for DSC, SEN, and PPV, respectively). Fig.6 shows that the CF$^2$-Net with super-pixel input can generate smoother segmentation results than the CF$^2$-Net without super-pixel input by providing textual similarity and contour integrality, which can utilize the textural information of the surrounding environment of the suspicious region to help determine the target region.

*4) Comparison with other methods*

TABLE IV
SEGMENTATION RESULTS FOR VALIDATING THE EFFECTIVENESS OF LOSS FUNCTION WITH WEIGHTED- BALANCED PARAMETER.

| Super-pixel | DSC(%) | SEN(%) | PPV(%) |
|---|---|---|---|
| × | 84.930±2.679 | 84.716±2.080 | 85.701±3.882 |
| √ | 85.553±1.718 | 85.211±1.342 | 88.198±1.988 |

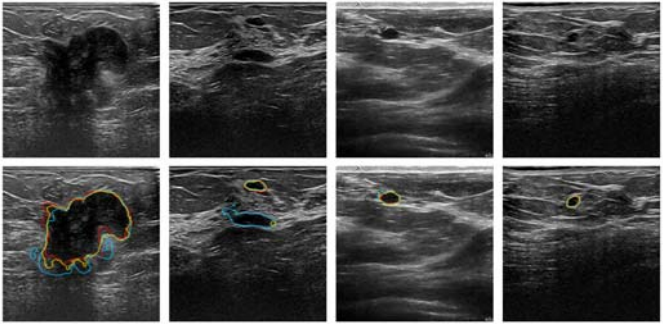

Fig.6. The visual comparison between the segmentation results of CF2-Net without/with super-pixel image as multi-input (blue/yellow), and the ground truth is also plotted (red).

We compare our proposed method with several state-of-the-art deep learning-based methods for BUS image segmentation, including U-Net, ConvEDNet, CE-Net, U-Net++, and Attention U-Net. These five competing methods are briefly introduced as follows.

• U-Net[30], which has been proposed to perform medical image segmentation with the architecture of contracting and expanding path. Notably, we have used it as the backbone of the CF2-Net and we have trained it by using our weighted-balanced loss function.

• ConvEDNet[32], which adopts a deep convolutional encoding–decoding network to address the difficult anatomical layer parsing problem and is armed with a deep boundary supervision module.

• CE-Net[39], which embeds a dense atrous convolution block and a residual multi-kernel pooling (RMP) block as context extractors in the middle of the encoding and decoding module.

• U-Net++[40], which essentially presents a deeply-supervised encoding–decoding network with a series of nested and dense skip pathways.

• Attention U-Net[41], which develops a novel attention gate (AG) model that automatically focuses on target structures with various shapes and sizes.

The comparison results are listed in Table V. The experimental results show that our proposed CF2-Net outperforms state-of-the-art deep learning-based methods for BUS image segmentation, which owe to three differences when compared with others, namely, i) a novel additional stream that hierarchically integrates symmetry shallow and deep features and captures additional edge information using an EC unit, ii) a weighted-balanced loss function with self-weight for each image to train the CF2-Net effectively, and iii) multi-channel inputs leverage super-pixel images to provide additional region. Both the CE-Net and the U-Net++ have achieved better results than the U-Net, because the RMP block of the CE-Net and the dense skip pathway of the U-Net++ can be considered as a contextual mechanism (also used in the PFS module of the CF2-Net) that can explore discriminant features. Then, we can also observe that the Attention U-Net is superior to the U-Net by embedding an attention mechanism that can automatically learn remarkable regions for segmentation, which can be the future work of the CF2-Net. In addition, the boundary-regularized ConvEDNet obtains poor performance, which may be because i) the used VGG-16 background is much deeper than the U-Net and have too many parameters to train, and ii) the strong boundary regularization is used in each scale of the encoding path, which is inclined to overfitting and ignores non-edged spatial information.

TABLE V
COMPARISION WITH DIFFERENT METHODS FOR BUS IMAGE SEGMENTATION.

| | DSC(%) | SEN(%) | PPV(%) |
|---|---|---|---|
| ConvEDNet | 73.709±1.809 | 79.459±2.617 | 72.983±1.447 |
| U-Net | 80.035±4.340 | 81.860±3.854 | 83.005±5.429 |
| CE-Net | 82.868±2.499 | 84.429±2.318 | 85.043±4.009 |
| Attention U-Net | 81.592±2.589 | 82.200±1.686 | 86.675±3.339 |
| U-Net++ | 81.301±1.459 | 80.452±1.617 | 88.186±3.534 |
| CF2-Net | **85.553±1.718** | **85.211±1.342** | **88.198±1.988** |

## IV. DISCUSSION AND CONCLUSION

In this paper, we have proposed a CF2-Net with the "E" type based on a novel feature integration strategy for BUS image segmentation. Specifically, the U-Net is used as the backbone network of the CF2-Net to generate coarse-to-fine feature maps. To further utilize the shallow features from encoder and deep features from decoder, we propose a novel FSP to integrate the shallow and deep information from the encoding and decoding path of the backbone network hierarchically, including ASPP units, CFF units, EC units, and tiny U-Net units to address the characteristics in the BUS image segmentation. In addition, to better segment lesion regions in a BUS image, we concatenate a super-pixel image and the original one as multi-channel inputs to release noise, enhance contour, and provide detailed structural. We also propose a weighted-balanced loss function with self-adaptive weight for each image to train the CF2-Net, which can relieve the imbalance issue between the target region and the background and adapt various sizes of the lesion region. The experimental results using four-fold validation demonstrated that the accuracy in the segmentation result of our proposed method are better than the state-of-the-art methods.

Although the proposed method has achieved the promising results in our experiments, it can be further improved. In future studies, we will improve our method from the following two aspects. First, in this work, the fusion of encoding and decoding features use the ASPP module to gain variously receptive fields but not to selectively learn the features. We will employ attention gates to select the information from the encoding and decoding before fusing them. Second, the location information provided by the super-pixel image is weak and priori. Thus, we will embed location constraints in the FSP module to capture the location feature in a data-driven way effectively.

In conclusion, we propose a new "E" type network, namely, the CF2-Net, for end-to-end BUS image segmentation. The novel fusion stream hierarchically integrates the shallow and deep information from the encoding and decoding path of the backbone network. In fact, the other useful constraints can also be embedded in the fusion stream to improve the network. The CF$^2$-Net (or "E" type network) is believed to be a general network and can be applied to other medical image segmentation tasks and it also is the be departure of 'E' type network.